\newcommand{\ket}[1]{| {#1} \rangle}

\documentclass[12pt]{iopart}
\usepackage{graphicx}

\def\m@thcombine#1#2{%
  \setbox0=\hbox{$#1$}
  \setbox1=\hbox{$#2$}
  \ifdim\wd0>\wd1
    \setbox0=\hbox to\wd1{\hss\box0\hss}
  \else
    \setbox1=\hbox to\wd0{\hss\box1\hss}
  \fi
  \mathop{\vcenter{
    \offinterlineskip\box0\box1}}}
\def\lesim{\m@thcombine<\sim}
\def\gesim{\m@thcombine>\sim}

\begin{document}

\title[Open problems in nuclear structure near drip lines]{
Open problems in nuclear structure near drip lines
}

\author{Masayuki Matsuo$^{1,2,3}$ and Takashi Nakatsukasa$^{3,2}$}

\address{$^1$Department of Physics, Faculty of Science, Niigata University, Japan}
\address{$^2$Graduate School of Science and Technology, Niigata University, Japan}
\address{$^3$RIKEN Nishina Center, Wako-shi, Saitama 351-0198, Japan
}
\begin{abstract}
Physics related to weakly bound nuclei and low-density asymmetric
nuclear matter
is discussed from a theoretical point of view.
Especially, we focus our discussion on new correlations in nuclei near
the drip lines and on open issues of the density functional theory
with a proper account of the continuum.
\end{abstract}

\maketitle

\section{Introduction}
\label{sec: intro}

Theoretical determination of the drip lines is one of 
the most difficult yet very important issues
in research in nuclear structure.
The existence of about 3,000 isotopes have been confirmed by experiments so far.
Still, up to three times as many are thought to be awaiting discovery.
The proton drip line is relatively easy to determine, because the electric
repulsion restricts the number of protons
accommodated in the nucleus with a given number of neutrons.
However, the location of the neutron drip line is practically {\it unknown}.
Experimentally, the neutron drip line is established only up to oxygen ($Z=8$).
Theoretically, it is even less!
Most nuclear models fail to reproduce the heaviest oxygen isotope.
Identification of the precise location of the neutron drip line will
continue to be a big challenge both for nuclear theory and experiment.

The stable nuclei are well characterized by
the basic property of nuclear systems,
known as the nuclear saturation \cite{BM69}.
It tells us, first, that
the binding energy per nucleon, $B/A\approx 8$ MeV, is approximately constant.
In stable nuclei, this is roughly equal to the Fermi energy
and to the separation energy (chemical potential) too,
$B/A\approx |\epsilon_F^{n(p)}| \approx S_{n(p)}$.
As the neutron number increases,
$B/A$ still stays roughly constant in neutron-rich nuclei.
However, $|\epsilon_F|$ and $S_n$ show a significant
decrease at specific neutron numbers (shell effects),
then, eventually disappear at the drip line.
Thus, to predict the neutron drip line, we need to calculate the
total binding energy $B$ (not $B/A$) in the precision of vanishing $S_n$.
A small error in calculation of correlation energy easily changes
the location of the neutron drip line by several units.

The nuclear saturation also says that
the nucleon density (near the center of the nucleus) is roughly constant
irrespective of the mass number.
The particle density,
which determines the Fermi momentum,
provides a typical energy (length) scale for fermion systems.
The phase-shift analysis of the nucleon-nucleon scattering
in the $^1S$ channel shows an attractive nature at low energy ($E<200$ MeV),
but becomes repulsive at high energy ($E>300$ MeV).
Therefore, the nuclear correlation is expected to change its character
according to the nuclear density.
In the ground states of stable nuclei with $\rho_n\approx\rho_p$ and
with $\epsilon_F^n\approx \epsilon_F^p$,
we are probing the nuclear correlation in a specific momentum (energy) range,
and the many-body correlations are somewhat ``screened''.
This situation is expected to change near the drip line,
such as nuclei with a neutron halo and a skin.
In addition, the loosely bound system with $\epsilon_F\approx 0$
will lead to spontaneous emission of nucleons from an excited level.
Therefore, all the excited states are basically embedded in the continuum
and require a proper treatment of the asymptotic behavior of the
many-body wave functions.

In this article, we present our personal view on current open issues
in nuclear theory at the drip lines.
Especially, we focus our discussion on theoretical account of
many-body correlations at low densities and in the continuum.

\begin{figure}[tb]
\begin{center}
\includegraphics[width=0.60\textwidth]{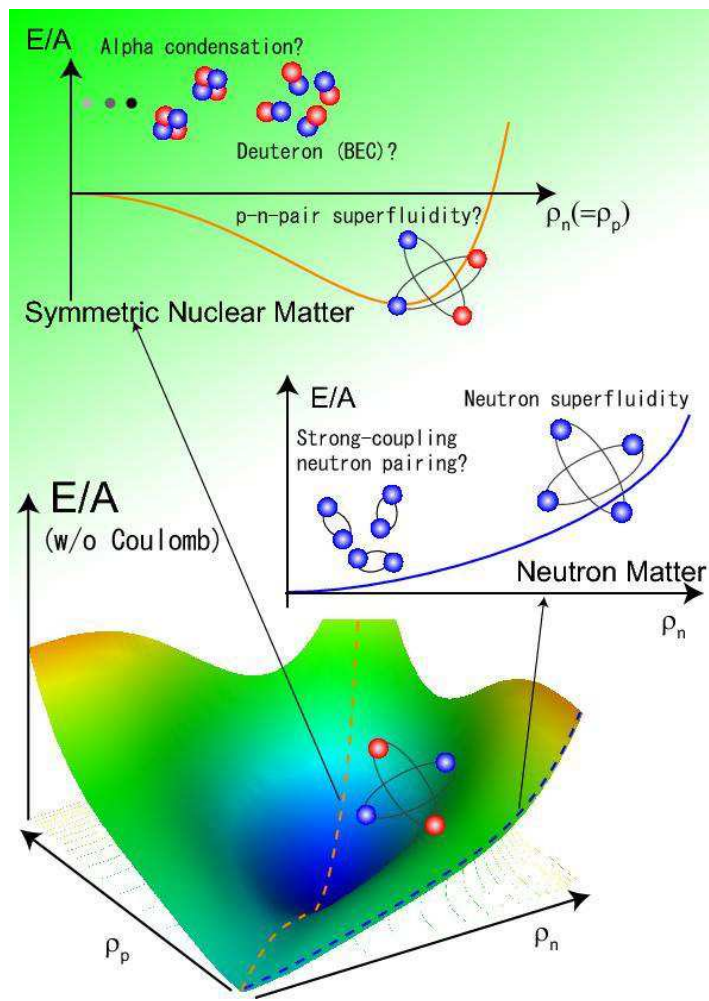}
\caption{{\small
Energy per nucleon of uniform nuclear matter calculated with
the Skyrme functional (SLy4),
as a function of neutron and proton densities, $\rho_n$ and $\rho_p$.
At low density, a variety of many-body correlations are expected to
manifest their presence.
}}
\label{fig:EOS}
\end{center}
\end{figure}

\section{Novel correlations in dilute matter and at drip lines}

\subsection{Correlations in dilute nuclear matter}
\label{sec: dilute_matter}

A nucleus, at the zeroth-order approximation, can be regarded
as a liquid droplet of nuclear matter.
Nuclear matter has
the nucleon density equilibrated at $\rho_0 \approx 0.17$ fm$^{-3}$ and
at the minimal asymmetry $\rho_n=\rho_p$ $(N=Z)$ without the
electromagnetic interaction.
The discovery of neutron halo and skin, however, indicates a
new possibility that the neutron-rich nuclei near the drip line
may not follow the conventional liquid-droplet picture.
They are provided with neutron-dominated density distributions
outside the (half-density) nuclear surface.
The thickness of the neutron skin reaches 
around 1 fm and the typical neutron density in the skin is
$\rho/\rho_0 \sim 10^{-1}$.
The neutron halo, whose size can be 
an order of several fm and even more, is characterized by
very low neutron density $\rho/\rho_0 \lesim 10^{-2}$ \cite{Ozawa01}. 
Clearly, the classical concept of the nuclear saturation
does not hold for these nuclei.

Is it possible to extract information on dilute neutron matter
from the observed properties of the neutron halo and skin?
Can we obtain the symmetry energy of asymmetric nuclear matter?
How significantly are these matter properties manifested by finite nuclei?
These are still open questions to answer.
Nevertheless, the analogy to dilute matter brings us a plenty of
inspiration on what physics can be anticipated in nuclei near the neutron
drip line. 

In fermion systems, it is generally expected that the correlations become
more eminent at relatively lower density, because the Pauli exclusion
effect is weakened.
New types of correlation are theoretically predicted,
some of which are shown in Fig. \ref{fig:EOS}.
For instance, the nuclear superfluidity, which has been known for a long time
in nuclear structure, is now expected to show a different character in the
neutron-rich environments at low density; crossover from the weak coupling
to the strong-coupling nature. 
We remark here that the pair correlation in general has two extreme limits
\cite{Leggett1,Leggett2}:
the {\it weak coupling limit} where the coherence length $\xi$
(the size of the Cooper pair)
is much larger than the average inter-particle distance $d$,
$\xi \gg d$, and the opposite
{\it strong coupling limit} where $\xi \ll d$.
The former corresponds to the pair correlation of electrons
in the standard metal superconductivity, while the latter to the Bose-Einstein
condensation (BEC) of strongly bound pairs of fermions.
Let us consider the pair correlation in 
a dilute gas of fermions with spin $1/2$.
The system is characterized by two parameters,
the Fermi momentum $k_F$ and
the scattering length $a$ in the $^{1}S$ channel.
The
weak and strong coupling limits are realized for $1/k_Fa \ll -1$ and $1/k_Fa \gg 1$,
respectively, and a crossover regime, characterized by
Cooper pairs whose size is comparable to
the average inter-particle distance ($\xi \sim d$),
shows up in a range $-1 \lesim 1/k_Fa \lesim 1$ \cite{Engelbrecht}.
Particularly, the middle point in the crossover, defined by $k_Fa =\pm \infty$,
is called the unitarity limit.  

The nuclear force in the $^{1}S$ channel is characterized, at the low energy,
by the large scattering length, $a \approx -20$ fm,
although the force strength 
strongly depends on the
relative momentum of interacting nucleons,
being less attractive with increasing momentum.
If we characterize the momentum dependence of the nuclear force by the 
effective range $r_e\approx 3$ fm,
dilute neutron matter is predicted to exhibit the signature
of the crossover ($\xi \sim d$)
in a range of $1/|a| \lesim k_F \lesim 1/r_e$,
which corresponds to a rather wide range 
of the neutron density, $10^{-4} \lesim \rho/\rho_o \lesim 10^{-2}$.
This estimate is indeed checked, in the zeroth order, 
with microscopic calculations performed 
at the levels of the Bogoliubov's
mean-field approximation \cite{Matsuo06}
or the one including higher-order medium effects \cite{Margueron}. 

The strong coupling pair correlation is predicted also
for the isoscalar ($T=0$) neutron-proton pairing in dilute symmetric matter.
In this case, even the ``deuteron condensation''
may be realized in some density range \cite{Alm93,Baldo95}.
In dilute symmetric matter,  a
{\it quarteting} correlation of two neutrons and two protons
may also play an important role.
This quarteting correlation may 
compete with the deuteron ($T=0$) correlation,
but might possibly lead to the ``alpha-particle condensation'' \cite{Beyer}.
It should be noted that the definitive answers
on these issues
are still to be awaited in future, especially for the
correlations in asymmetric nuclear matter.

It is also a non-trivial and open question whether these correlations
characteristic 
to dilute matter can be found in finite nuclei. There exist some
symptoms: A pair of weakly bound neutrons in two-neutron-halo nuclei has been
predicted to exhibit a spatial di-neutron correlation \cite{Bertsch91,Zhukov}.
Recent experimental investigations
on $^{11}$Li and $^{6}$He support these predictions \cite{Nakamura,Wang}. 
Recent Hartree-Fock-Bogoliubov mean-field calculations performed for  
medium- and heavy-mass nuclei suggest that
a significant di-neutron correlation 
emerges rather generically in the surface of nuclei, especially in neutron-rich
isotopes \cite{Matsuo05,Pillet}.
The famous Hoyle state ($0_2^+$) in $^{12}$C, which was previously studied with
a cluster model of three alpha particles, has been
revisited and is shown to exhibit a low-density profile
with three alpha particles ``condensed'' in a single $s$-orbit of
the relative motion \cite{THSR}.
It is of crucial importance to investigate possible
link between these observations and the exotic correlations
in dilute nuclear (neutron) matter.

\subsection{Correlations in weakly bound nuclei}

The single-particle motion in the nucleus is {\it quasi-stable},
since the mean-free path of nucleons is larger than typical nuclear size,
$\lambda_{\rm mf} \gg R$.
In other words, the collision time is larger than the typical
time scale of nucleonic motion,
$\tau_{\rm coll} \gg mR/(\hbar k_F)\sim 10^{-22}$ s.
This leads to the shell-model picture of the nucleus,
which consists of nucleons independently moving
in an average nuclear potential (self-trapped Fermi gas).
In this picture,  the neutron (proton) drip-line is defined by
the zero Fermi energy of neutrons (protons).
The single-particle orbitals at zero energy ($\epsilon\approx 0$)
are very different from those of stable nuclei with $\epsilon<$
several MeV.
Especially, the neutron orbital with small orbital angular momentum $l$
exhibits a long tail extending far
outside the nuclear surface.
This is nothing but the origin of the neutron halo,
which has been understood by the tunneling into
the classical forbidden region.
Can these neutrons travel without the collision longer than stable nuclei?
We need a microscopic picture of the nucleus at the drip lines,
with full account of short-range nucleon-nucleon correlation and
long-range many-body correlation in the continuum \cite{DB04,MNP07}.
Does the paring correlation suppress the halo effect (anti-halo effect)
\cite{BDP00,HM03}?
Furthermore, the three-body interaction might also play an important role
at the drip line.
In fact, a recent work suggests that the drip line of the oxygen isotopes
is significantly affected by the Fujita-Miyazawa-type three-body interaction
\cite{Otsuka09}.

A nucleus at the neutron drip line in the medium- and heavy-mass region
may have several loosely bound orbitals
which are able to accommodate many neutrons.
What kind of many-body correlations and collective modes of excitation
can we expect in such circumstances?
In stable nuclei, the collective motion at low energy has mostly
the isoscalar character in which the neutrons and protons
move together.
In the drip-line nuclei with highly asymmetric $N/Z$ ratio,
the protons and neutrons have very different Fermi momenta.
Does this lead to independent collective motion of neutrons and protons?

An example of the excitation profiles peculiar to drip-line nuclei
is given by the large Coulomb dissociation 
cross section leading to emission of neutrons.
In the case where 
essentially only one valence nucleon is loosely bound,
such as a typical one-neutron-halo nucleus $^{11}$Be \cite{NakamuraBe11,Aumann},
the large electric dipole ($E1$)
strength just above the neutron separation energy
can be interpreted in terms of a simple
mechanism of the single-particle
transition from a weakly bound orbital to low-energy continuum orbitals.
In nuclei containing two or more loosely bound nucleons,
however, 
the situation is not that simple as 
the correlation among these nucleons comes into the game.
It has been discussed for a long time,
but the correlation in the Coulomb dissociation process
is still an open issue  for two-neutron-halo nuclei
\cite{Bertsch91,Zhukov,Myo}.

Nuclei with more than two loosely bound nucleons
are appropriate to be called weakly-bound many-body systems.
An example may be the ``next'' drip-line nuclei   
in the $Z = 10\sim 20$ (Ne-Ca) region.
A density-functional calculation
predicts occurrence of the shape deformation \cite{Sto03}.
For this kind of prediction, the precise calculation of the one-body
(mean-field) potential with proper account of the continuum 
single-particle wave function and the pair correlation
is necessary over the entire nuclear chart.
The presence of loosely bound nucleons has a feedback to the one-body
potential, which may lead to
the shell evolution and change of magic numbers.
The shell structure is essential for
determination of nuclear deformation.
Since weakly bound and unbound neutrons are now combined with the deformation
and presumably with the strong pair correlation, 
they will then bring about new features in the low-energy $E1$ mode
and possibly in other modes of excitation.


In heavier mass region, the low-energy electric dipole peaks (pygmy
resonances) are observed systematically in many nuclei including stable ones
\cite{Aumann,KPZ06}.
The character of the pygmy resonance may vary
as we increase $N$ along an isotopic chain, i.e., with decrease of
the neutron separation energy, and furthermore it may also depends on the
mass regions.
How collective are the pygmy dipole resonance? What is the mechanism
of causing the pygmy resonance?  How the pygmy dipole resonance is
connected to the low-energy dipole mode which emerges in drip-line 
containing very weakly bound neutrons.
We have not reached a unique answer on the nature of these
pygmy excitations.
Some calculations suggest a decoupled vibrational motion of the neutron skin,
but some suggest single-particle excitation of loosely bound neutrons
\cite{PVKC07}.
The influence of the strong pairing and the di-neutron correlation expected
in the low-density neutron skin
is also pointed out
\cite{Matsuo05}. 
To reach the conclusion, again, we need a large-scale calculation with full
account of the correlations and the continuum.
It is certainly an open problem
for nuclear structure theory to make a reliable theoretical 
prediction and precise description of the excitation modes.
Note that the properties of the pygmy resonances has a significant impact
on the neutron capture cross section, one of the most important reactions
for the element synthesis in the universe.

We believe that the nuclear density-functional approaches
including their dynamical
extensions are currently
one of the most promising theoretical framework
which can
incorporate in a unified way many of the features mentioned above:
>From the weak coupling features at normal nuclear density $\rho\approx\rho_0$
to the strong coupling aspects in dilute nuclear matter, in the nuclear
surface and exterior regions of neutron-rich nuclei.
The time-dependent density functional theory may, in principle, describe
elementary modes of excitation associated with 
the loosely bound and unbound nucleons, from stable to drip-line nuclei,
with the universal functional \cite{RG84}.
Let us describe our view of open problems in the density functional approaches
in the next section.

\section{Nuclear density functional theory and
 open problems at drip lines}

\subsection{Energy density functional}

The universal energy density functional, $E[\rho]$,
is a key of the density functional theory \cite{LPT03,Sto03}.
Despite of recent efforts to improve the energy density functional,
it is still far from solid prediction of the drip lines
(see Sec.~\ref{sec: intro}).
A major problem at the proton drip line is related to the Wigner
energy in $N=Z$ nuclei.
It cannot be explained by conventional density functionals and
its extension is necessary.
There are some arguments that the isoscalar ($T=0$) pairing might
be responsible for the Wigner energy \cite{SW97,Nee02},
but it is not conclusive yet.

The density functional for electronic many-body systems is constructed
so as to reproduce high-density and low-density limits for which the
exact answers are known either analytically or numerically.
This is not the case in nuclear systems.
The current nuclear theory still lacks of a desirable knowledge about
properties of nuclear matter.
The exact solution
is unknown even at these low- and high-density limits, because of
nuclear clustering at low density, and lack of knowledge about
the three-body (many-body) interactions
that will become important at high density.
Thus, in the nuclear systems, empirical information is crucial
for determination of the functional form.
Since the empirical data on stable nuclei are restricted
near the nuclear saturation density, $\rho\approx\rho_0$,
data near the drip lines provide indispensable information about
the nuclear density functional.

In contrast to a number of difficulties in nuclear matter,
pure neutron matter is relatively easier to study theoretically.
Properties of high-density neutron matter,
that is important for physics of neutron stars,
are not conclusive yet, due to ambiguities in the three-body interaction.
However, low-density neutron matter
can be studied by precise many-body calculations with a two-body
interaction determined by the phase-shift analysis.
As is discussed in Sec.~\ref{sec: dilute_matter},
one of the important features in low-density neutron matter is
a strong-coupling nature of the pair correlation.
Using the Bogoliubov's quasiparticle formalism,
the pairing energy can be taken into account with a functional 
of the pair (abnormal) density,
$\tilde\rho(\vec{r})
=\langle \psi(\vec{r}\tilde\sigma)\psi(\vec{r}\sigma)\rangle$.
Its functional form,
$E_{\rm pair}[\rho_n,\rho_p,\tilde\rho_n,\tilde\rho_p]$,
is then need to be determined, but it is an open problem currently under
discussion \cite{DNS02,MSH07,YSN09}.
We need an elaborate pairing functional $E_{\rm pair}[\rho_n,\rho_p,\tilde\rho_n,\tilde\rho_p]$
which can cover from the low density limits to the saturation density
in order to describe physics at the drip lines.

\subsection{Many-particle dynamics in the continuum for nuclei at the drip lines}

Many of nuclear structure calculations utilize the harmonic oscillator
basis functions.
It is very efficient for deeply-bound nuclei.
However, it is becoming less reliable near the drip lines, especially
at the neutron-rich side.
The continuum particularly plays an important role in dynamical aspects
of nuclei near the neutron drip line.
Excitation functions of weakly-bound systems have continuous spectra
and resonant states are embedded in the sea of background continuum.
Then, the validity of the $L^2$ basis-set approximation
that utilizes the discrete basis functions comes into question.

For unbound (positive-energy) states,
the theoretical treatment of the continuum requires a proper 
boundary condition of many-body wave functions.
This problem becomes non-trivial even for there-body systems;
for instance, a notorious scattering problem with three charged particles.
The boundary condition is required when we solve the stationary quantum
mechanical problem.
For scattering states, the solution can be formally
expressed by the M\o ller's wave operator as
\begin{equation}
\label{scattering_state}
\fl
\ket{\Psi^{(+)}(E)}= U(0,-\infty) \ket{\Phi(E)} 
                   =\left( 1-\int_{-\infty}^0 dt' U(0,t')
                     e^{-\eta|t'|/\hbar}V_I(t') \right) \ket{\Phi(E)} ,
\end{equation}
where $U(t,t')=e^{iH_0 t/\hbar}e^{-iH(t-t')/\hbar}e^{-iH_0 t'/\hbar}$ and
$V_I(t)\equiv e^{iH_0 t/\hbar}V e^{-iH_0 t/\hbar}$.
Here, the Hamiltonian is divided into two parts, $H(t)=H_0+Ve^{-\eta |t|/\hbar}$,
and the initial state, $\ket{\Phi(E)}$,
is the energy eigenstate of $H_0$.
We need the convergence factor $e^{-\eta|t'|/\hbar}$,
to remove the interaction between a projectile and a target
at the asymptotic limit of $t\rightarrow -\infty$.

Time-dependent methods may offer a number of advantages over the stationary
methods, especially with respect to the treatment of the continuum.
This can be demonstrated for the solution of scattering state in Eq.
(\ref{scattering_state}).
A single time propagation starting from a wave packet,
$\ket{\Phi_{\rm WP}}=\ket{\Psi(t=-\infty)}$
instead of the energy eigenstate $\ket{\Phi(E)}$,
provides the $S$ matrix for a broad energy interval.
Since the wave packet is spatially localized,
we can set the initial state so as to satisfy the condition,
$V\ket{\Phi_{\rm WP}}=0$,
for the finite-range interaction.
Therefore, the infinitesimal parameter $\eta$ becomes unnecessary and
the proper boundary condition is automatically imposed.
In addition, the time-dependent methods are
relatively easy to program and suitable for the parallel computing.
The power of the time-dependent method has been testified by
its applications to fusion dynamics of halo nuclei, for which
the three-body Schr\"odinger equations need to be solved exactly
\cite{IYNU06}. The method
is also applied to calculation of the fully-self-consistent dynamical
response in the full three-dimensional coordinate-space representation
using the time-dependent density functional theory \cite{NY05,NY01}.

Developments of the time-dependent description of
many-particle dynamics are heading for several directions.
Firstly, we need to
include the pair correlation in the density-functional approaches,
which is essential for description of weakly-bound nuclei in medium- and
heavy-mass regions. In other words, we should solve the time-dependent
Hartree-Fock-Bogoliubov (TDHFB) equations which can describe the dynamics of
the pair correlated many-body systems on the basis of
the pairing functional
$E_{\rm pair}[\rho_n,\rho_p,\tilde\rho_n,\tilde\rho_p]$. This extension is essential in order
to widen the field of application, including drip-line nuclei in medium- and heavy-mass
regions, and hopefully to describe two-proton decay from nuclei near the
proton drip line. 
Another current effort is
the time-dependent description of nuclear reaction with
more than three particles.
We anticipate that massive parallel computers may allow us to
attack this subject.
For reactions with even more degrees of freedom such as
heavy-ion reactions,
we can exploit the time-dependent density functional theory
as it provides, in principle,
an {\it exact} description of the nuclear dynamics \cite{RG84}.
However, calculation of the exclusive cross sections
have some conceptual open problems we need to solve.
This is similar to the famous problem of spurious cross channel correlations
in TDHF \cite{RS80}.
Some hints have been given by the path integral formulation of the TDHF
\cite{Rei82}, however, there still remains a long way to achieve
the final goal.

\subsection{Linear response calculations: elementary modes of excitation
 at the drip lines}

\begin{figure}[tb]
\begin{center}
\includegraphics[width=0.60\textwidth]{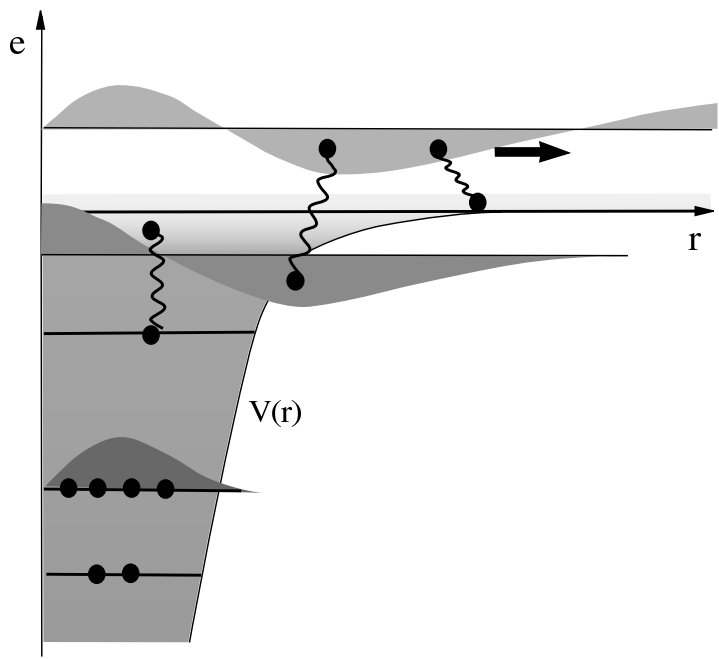}
\caption{{\small
Schematic picture depicting correlations in
nuclei near the drip-line, for which correlations involving 
the loosely bound and
unbound continuum orbits play essential roles.
}}
\label{fig:correlations}
\end{center}
\end{figure}

In order to understand complex many-body systems such as nucleus,
the concept of collective motion (collective degrees of freedom)
is often very useful.
Elementary excitations, small fluctuations around the ground state
of many-particle systems, are basic building blocks of the collective motion, 
and they
simplify and clarify characteristic features of
 excitation spectra. 
Search for the new elementary excitations  is therefore one of
the most important subjects
in the studies of nuclei near the drip-line (see. Section 2.2).

The time-dependent density functional theory provides us
the best framework to describe the elementary excitations.  Once a stationary equilibrium solution
is constructed from the variational principle of the density functional, 
we then like to describe the linear response of the system against external perturbations
using the consistent functional. 
The linear regime of the time-evolution is equivalent to the
random phase approximation (RPA).
Thanks to the wide applicability of the density-functional
theory, we expect to treat various situations in a unified framework: from
stable to drip-line nuclei, both spherical and deformed configurations, both 
weak and strong coupling pair correlations, 
and from neutron matter to inhomogeneous nuclear matter such as the inner crust
of neutron stars. In order to enable the unified description, however,
the methodology of the linear response theory and
the RPA needs to be developed. Treatment of
the continuum is of special importance for nuclei near the drip-line
(Fig. \ref{fig:correlations}).

The linear response theory based on the density functional theory,
including the continuum
and any kind of possible shape deformations,
has already been given a formulation \cite{NY05,NY01}.
A big open problem is, therefore,
to develop the quasiparticle random phase approximation (QRPA) techniques which
can include both the pair correlation and the continuum on top of the features mentioned above. 
This is nothing but to formulate a linear response theory on the basis of the
time-dependent Hartree-Fock-Bogoliubov (TDHFB) equations for the time-evolution
of the pair correlated systems.  A difficult part of the many-body theory is 
that in the Bogoliubov's
quasiparticle scheme there is no distinction between the hole and the particle
orbitals as they couple through the pair potential.
The number of occupied orbitals
we deal with in the RPA is just the number of nucleons, $A$, while in the
QRPA, it is equal to the total number of quasiparticle orbitals
in the whole phase
space, and it is ``infinity'' if we explicitly treat the continuum orbitals.
The Green's function technique circumvents this problem
\cite{Matsuo01,KMS09,OM09}, but
extensions to deformed drip-line nuclei are still being awaited.
The QRPA calculation including all the residual interactions (fields)
in the density functional is another direction in pursuit.
Spherical codes \cite{Paar,Terasaki} are available already, and
the current R\&D's are focused on the case of the axially symmetric deformation
\cite{PR08,PG08,YG08}.
The finite amplitude method \cite{NIY07,INY09}
may provide a feasible alternative approach for this purpose.
The linear response calculation with full self-consistency and the continuum
requires a large-scale
numerical effort, for which we need to develop a new methodology and
an algorithm suitable for
the massive parallel computation. Ultimately we hope to achieve 
the linear response theory applicable not only to finite nuclei but also
inhomogeneous nuclear matter, for which a full techniques of
condensed-matter physics better to be imported. 

Finally we mention that 
the static and the linear response formulations of
the density functional theory should be regarded as a ground base for
further developments. 
It can be extended further to describe 
the large-amplitude non-linear dynamics
(See Ref.~\cite{MMNHS10}), restoration of the symmetries, and 
higher-order many-body effects such as exchanges of elementary modes.
In the end, we hope to understand the nuclear many-body systems
from a unified picture.

\section{Summary}

Nuclei near the drip lines open new possibilities for research in
nuclear physics.
Nuclei in new circumstances, such as
weakly-bound many-body systems with
high asymmetry at low density,
we may expect new types of correlation, different from our conventional
picture of the nucleus.
Simultaneously, we are now facing interesting but 
difficult tasks to solve basic issues of nuclear many-body problems.
In this article, we did not intend a broad review of open theoretical
issues related to the drip lines.
Instead, we focused our discussion on open problems associated with
the continuum and the nuclear density functional approaches.
Currently, the nuclear structure theory
is entering a revolutionary stage,
which demands us to
construct microscopic models with
precision much higher than it used to be.

\ack

This work is supported by the Grant-in-Aid for Scientific Research
(Nos. 20540259, 21105507, 21340073 and 20105003).

\clearpage

\section*{References}

\end{document}